\documentclass[12pt]{amsart}
\usepackage{amsmath,amssymb}
\usepackage{hyperref}
\usepackage{wasysym} 
\usepackage{graphicx}
\usepackage{mathtools}
\usepackage{cite}
\usepackage[foot]{amsaddr}
\usepackage{bussproofs}
\usepackage{tikz-cd}

\newtheorem{thm}{Theorem}
\newtheorem{cor}{Corollary}
\newtheorem{lem}{Lemma}
\newtheorem{prop}{Proposition}

\newtheorem{eg}[thm]{Example}

\newtheorem{defn}{Definition}

\newtheorem{rem}[thm]{Remark}

\DeclarePairedDelimiter{\norm}{\lVert}{\rVert}
\newcommand{\badstart}[0]{\ \\[-.2in]}
\title{An Inductive Construction for Many-Valued Coalgebraic Modal Logic}

\date{}

\author{CHUN-YU LIN and CHURN-JUNG LIAU}

\address{Institute of Information Science
Academia Sinica, Nankang 115, Taipei, Taiwan}

\email{liaucj@iis.sinica.edu.tw.}

\begin{document}

\begin{abstract}
In this paper, we present an abstract framework of many-valued modal logic with the interpretation of atomic propositions and modal operators as predicate lifting over coalgebras for an endofunctor on the category of sets. It generalizes Pattinson's stratification method for colagebraic modal logic to the many-valued setting. In contrast to standard techniques of canonical model construction and filtration, this method employs an induction principle to prove the soundness, completeness, and finite model property of the logics. As a consequence, we can lift a restriction on the previous approach~ \cite{Lin2022} that requires the underlying language must have the expressive power to internalize the meta-level truth valuation operations.
\end{abstract}
\maketitle
\section{Introduction}\label{sec1}
Many-valued logic and modal logic are both long-standing themes in the research of symbolic logic and its applications since the last century. While the former is the main formalism for reasoning about vague information in terms of (mathematical) fuzzy logic~\cite{cintula2011handbook}, the latter has found diverse applications in software engineering, AI, economics, and philosophy~\cite{hml}.  In past decades, there have been increasing interests in systems integrating these two logics. From the early work motivated by reasoning with incomplete information~\cite{Ostermann88,Ostermann90} and multi-expert opinions~\cite{fitting1991many,fitting1992many} to more recent studies on mathematical fuzzy logic~\cite{bou2011minimum,fml10,fml12,fml13,lmcs11
}, such integration not only extends application scopes of the respective formalism, but also raises interesting theoretical issues.

Nowadays, there are many different variants of modal logic  because of its extensive application scopes. Hence, it is desirable to have a general framework to encompass a variety of semantics of modal logics. Coalgebraic modal logic, first proposed by Moss~\cite{moss1999coalgebraic}, provides such a framework. There are two main approaches to coalgebraic modal logic-the relation lifting\cite{moss1999coalgebraic} and the predicate lifting\cite{pattinson2003coalgebraic,schroder2007finite}. Venema et al. first studied a sound and complete axiomatization for coalgebraic modal logic through relation lifting\cite{venema2012completeness}. On the other hand, the soundness and completeness of coalgebraic modal logic under predicate lifting were investigated in  \cite{schroder2007finite} and \cite{pattinson2003coalgebraic} using finite-model constructions and induction on modal ranks respectively. For a more comprehensive survey of coalgebraic modal logic, see \cite{kupke2011coalgebraic}.

Analogously, many-valued coalgebraic modal logic can provide a uniform framework for a number of many-valued modal logics. Along this direction, B{\'\i}lkov{\'a} and  Dost{\'a}l studied both relation lifting under quantales\cite{bilkova2013many}  and predicate lifting under finite residuated lattices \cite{bilkova2016expressivity}. With some extra assumptions on the predicate lifting or relation lifting, they show that many-valued coalgebraic modal logic has Hennessy-Milner property. Also, Schr\"oder and Pattinson developed a coalgebraic semantics over standard {\L}ukasiewicz algebra for fuzzy description logics and fuzzy probabilistic logics\cite{SchroderP11}. A well-known result in coalgebraic modal logic is that its soundness and completeness can be determined at the one-step level. In \cite{Lin2022}, it is shown that the same result also hold in the many-valued case by using the technique of canonical model construction developed in \cite{schroder2007finite,kupke2015weak}. However, when applied to the many-valued case, the construction requires that the language must have the expressive power of internalizing the meta-level truth valuation operations. To lift that restriction, we adopt the induction method proposed in \cite{pattinson2003coalgebraic} to prove the soundness and completeness of many-valued coalgebraic modal logics. In addition, we can prove the finite model property under the inductive construction without using filtration by imposing the finiteness assumption on the functor. This gives rise to an alternative approach to study meta-properties of many-valued coalgebraic modal logics.

This paper is structured as follows. In the next section, we introduce the preliminary concepts and notations used in this paper. In Section~\ref{sec3}, we present the syntax and semantics of many-valued coalgebraic modal logic $\mathcal{ML}$ and its rank $n$ fragement $\mathcal{L}_n$, and show how to connect them using the technique of induction along the terminal sequence developed in \cite{pattinson2003coalgebraic}. In Section~\ref{sec4}, we define proof systems of $\mathcal{ML}$ and $\mathcal{L}_n$ as consequence relations $\mathbf{L}$ and $\mathbf{L}_n$ respectively, and also establish the connection between them.  In Section~\ref{sec5}, we present one-step logic in the many-valued setting and demonstrate how to prove soundness and completeness of $\mathbf{L}$ by assuming its one-step soundness and completeness. Finally, we prove the finite model property assuming the finiteness of the underlying functor in Section~\ref{sec6} and conclude the paper in Section~\ref{sec7}. In addition, we include basic notions of category theory and several instances of the many-valued coalgebraic modal logic in the appendices. 

\section{Preliminaries and Notations}\label{sec2}
In many-valued logic, we usually generalized the set of truth values from the Boolean algebra $\mathbf{2}$ to an lattice $\mathbb{A}$. The many-valued structure considered in this paper is the (finite) residuated lattice, which provides semantics for a wide class of substructural logics~\cite{Ono2003}.
\begin{defn}
We say that $\mathbb{A}= \langle A,\lor^{\mathbb{A}},\land^{\mathbb{A}},\to^{\mathbb{A}},\odot^{\mathbb{A}},0^{\mathbb{A}},1^{\mathbb{A}} \rangle$ is a {\it commutative integral Full-Lambek algebra(FL-algebra) } (aka residuated lattice) if
\begin{itemize}
    \item $\langle A,\lor^{\mathbb{A}},\land^{\mathbb{A}},0^{\mathbb{A}},1^{\mathbb{A}} \rangle$ is a bounded lattice,
    \item $\langle A, \odot^{\mathbb{A}}, 1^{\mathbb{A}} \rangle$ is a commutative monoid,
    \item We can define ordering $\leq^{\mathbb{A}}$ as $a\leq^{\mathbb{A}} b$ iff $a \land^{\mathbb{A}} b=b$ iff $a \lor^{\mathbb{A}} b =a$,
    \item  $\to^{\mathbb{A}}$ is the  residuated implication with respect to $\odot^{\mathbb{A}}$, i.e. for all $a,b,c \in A$, $a\odot^{\mathbb{A}} b \leq^{\mathbb{A}} c$ iff $b \leq^{\mathbb{A}} a \to^{\mathbb{A}} c$,
    \item $a \leq^{\mathbb{A}} 1^{\mathbb{A}}$ for all $a \in A$.
\end{itemize}
\end{defn}
We will omit the superscript for each operation on the FL-algebras $\mathbb{A}$ without causing any confusion. When we mention an FL-algebras $\mathbb{A}$, we always use $A$ to denote its carrier set.

To introduce the notion of predicate lifting, we assume the familiarity of basic category theory, mainly the definitions of category, functor, and natural transformation~\cite{Awodey}. In this paper, we are exclusively concerned with the category of sets, denoted by $\mathbf{Set}$, whose objects are sets and morphisms are functions between sets. We assume that $T: \mathbf{Set}\to \mathbf{Set}$ is a  nontrivial endofunctor, i.e. there exists a set $S$ such that $TS\neq \emptyset$.
\begin{defn}
A $T$-coalgebra is a pair $\langle S,\sigma \rangle$ where $S$ is a set and $\sigma:S \to TS$ is a function.
\end{defn}

Recall that for any two sets $X$ and $Y$,  a Hom-set $Hom(X,Y)$ denote the set of all functions (i.e. morphisms) from $X$ to $Y$. In addition, $Hom(-,Y): \mathbf{Set}\to \mathbf{Set}$  is a contravariant functor that sends a set $X$ to $Hom(X,Y)$ and a function $f: X_1\to X_2$ to a function $Hom(f,Y):Hom(X_2,Y)\to Hom(X_1,Y)$ such that for any $g\in Hom(X_2,Y)$, $Hom(f,Y)(g)=g\circ f$. We also call $Hom(-,Y)$ a Hom-functor. We adapt the definition of predicate lifting proposed in \cite{bilkova2016expressivity}. For simplicity, we only consider the unary case in the rest of this paper. One can easily generalize all the results to $n$-ary cases for any $n \geq 1$.
\begin{defn}
A {\it predicate lifting} for $T$ is defined as a natural transformation $$\lambda: Hom(-,A) \Rightarrow Hom(T(-),A)$$ with FL-algebra $\mathbb{A}$.
\end{defn}
Because a natural transformation is a family of morphisms indexed by objects of the category, we use $\lambda_S$ to denote the morphism $$Hom(S,A)\to Hom(TS,A)$$ for an object $S$ in $\mathbf{Set}$.

In coalgebraic logic, we can regard propositional symbols as nullary modalities. To do that, we first define the product functor $T_P = EV_P\times T$ for a set of proposition symbols $P$, where $EV_P$ is a constant functor which maps an object (i.e.\ a set) to $Hom(P,A)$ and a morphism to $\text{id}_{Hom(P,A)}$ (i.e.\ the identity function from $Hom(P,A)$ to itself). The $T_P$-coalgebra is a pair $\langle S,\sigma_V\rangle$ where $S$ is a set and $\sigma_V:S\to T_PS$ is defined by $$\sigma_V(s)\coloneqq\langle V(s),\sigma(s)\rangle,$$
where $V:S\to Hom(P,A)$ is a valuation of propositional symbols over $S$ and $\sigma:S\to TS$ is a $T$-coalgebra. Then, for any propositional symbol $p\in P$, the nullary predicate lifting $$\lambda^p:Hom(-,A^0)\Rightarrow Hom(T_P(-),A)$$ for the functor $T_P$ is defined as  $$\lambda_S^p(!_S)(\nu,\delta)\coloneqq\nu(p)$$ for $\nu\in EV_P S$, $\delta \in TS$, and the unique map $!_S:S\to A^0$. Note that $A^0$ is a singleton set which is the terminal object in $\mathbf{Set}$. Also, we use the predicate lifting $\lambda$ to define $$\lambda':Hom(-,A) \Rightarrow Hom(T_P(-),A)$$ as  $$\lambda_S'(X)(\nu,\delta) \coloneqq\lambda_S(X)(\delta) $$ for any $X:S\to A$. From now on, we assume that $\Lambda$ is a set of (unary) predicate liftings and define $\Lambda_P$ as $\Lambda\cup\{\lambda^p| p\in P\}$.

\section{Syntax and Semantics}\label{sec3}
In this section, we present the syntax and semantics of many-valued coalgebraic modal logic.
\subsection{Syntax}
Given an FL-algebra $\mathbb{A}$, an endofunctor $T$ over $\mathbf{Set}$, a set of propositional symbols $P$, and a set of predicate liftings $\Lambda$, the alphabet of our language consists of logical connectives $\lor$, $\land$, $\&$, $\to$, modalities $ \varhexagon_{\lambda}$ and $\varhexagon_{\lambda^p}$ for every $\lambda\in\Lambda$ and $p\in P$, and constant symbols $\bar{c}$  for every $c\in A$ when $A$ is finite and $c\in\{0,1\}$ when $A$ is infinite. The language $\mathcal{ML}$ is then defined inductively as follows:$$  \varphi ::=\bar{c}\;|\;\varphi\lor\varphi\;|\; \varphi\land\varphi \;|\;\varphi\&\varphi\;|\;\varphi\to\varphi\;|\;  \varhexagon_{\lambda}\varphi \;|\;\varhexagon_{\lambda^p}.$$
We abbreviate $(\varphi\to\psi)\land(\psi\to \varphi)$ as $\varphi\leftrightarrow\psi$. Moreover, because we identify the propositional symbol $p$ with the nullary modality $\varhexagon_{\lambda^p}$, we usually write $p$ in place of $\varhexagon_{\lambda^p}$ in a formula.

\begin{rem}
We assume that there is a sound and complete many-valued logic $Ax(\mathbb{A})$ with respect to a FL-algebra $\mathbb{A}$. When $\mathbb{A}$ is a finite FL-algebra, such an $Ax(\mathbb{A})$ necessarily exists (see the appendix in \cite{bou2011minimum}). In this case, $\mathcal{ML}$ contains a constant symbol $\bar{c}$ for any $c\in A$. If $\mathbb{A}$ is infinite, $Ax(\mathbb{A})$ exists in some (e.g. standard BL, {\L}ukasiewicz, G$\ddot{o}$del, and product algebras; see \cite{cintula2011handbook}) but not necessarily all cases. For the infinite-valued case, $\mathcal{ML}$ only need to contain constants $\bar{0}$ and $\bar{1}$.
\end{rem}

To use the induction principle, we need to stratify $\mathcal{ML}$ according to modal ranks of formulas.
\begin{defn}
The rank of a formula $\varphi \in \mathcal{ML}$ is defined inductively as follows :
\begin{itemize}
    \item rank($\bar{c}$)=rank($\varhexagon_{\lambda^p}$)=0,
    \item rank($\varphi \ast \psi$) = max$\{\text{rank}(\varphi),\text{rank}(\psi)\}$ for $\ast\in\{\vee,\wedge,\&,\to \}$,
    \item rank($\varhexagon_{\lambda}\varphi$) = rank($\varphi$)+1
\end{itemize}
\end{defn}

We define the rank-$n$ language $\mathcal{L}_n$ for each $n\in \omega$ as follows. First, we define the rank-0 language $\mathcal{L}_0$ inductively as
$$\pi::= \;\bar{c}\;|\;\pi\lor\pi\;|\;  \pi\land\pi\;|\;\pi\&\pi\;|\;\pi\to\pi\;|\;\varhexagon_{\lambda^p}.$$
Second, for any nonempty set $\Phi$, let $\Lambda(\Phi)$ denote $$\{\varhexagon_{\lambda}\varphi|\lambda\in\Lambda,\varphi\in\Phi\}.$$ and let
$\mathcal{L}(\Phi)$ be the smallest set containing $\Phi$ and closed under the operation of $\lor$, $\land$, $\&$, and $\to$. Then, the rank-$n$ language is defined as $$\mathcal{L}_n=\mathcal{L}(\Lambda(\mathcal{L}_{n-1})\cup\mathcal{L}_{n-1})$$ for any $n \geq 1$. By construction, we can see that $\mathcal{ML}= \bigcup_{n \in \omega} \mathcal{L}_n$, where $\mathcal{L}_n$ contains all $\varphi \in \mathcal{ML}$ with rank$(\varphi) \leq n$.

\subsection{Semantics}
Next, we develop the semantic for $\mathcal{ML}$. There are two different models which correspond to the layered structure of $\mathcal{ML}$. For the full language $\mathcal{ML}$, the semantic is defined as follows.
\begin{defn}
A {\it T-model} for $\mathcal{ML}$ is a $T_P$-coalgebra $\mathfrak{C}= \langle S,\sigma_V\rangle$ consisting of a nonempty set of states $S$ and a map $\sigma_V:S\to T_PS$ defined by $\sigma_V(s)\coloneqq\langle V(s),\sigma(s)\rangle$, where $V:S\to Hom (P,A)$ is a valuation of propositional symbols $P$ over $S$ and $\sigma:S\to TS$ is a $T$-coalgebra. We define the semantics $\norm{\varphi}_{\mathfrak{C}}:S\to A$ for $\varphi\in\mathcal{ML}$ inductively such that for all $s\in S$
\begin{itemize}
    \item $\norm{\bar{c}}_{\mathfrak{C}}(s) \coloneqq c$ with $c\in A$,
    \item $\norm{\varphi\ast\psi}_{\mathfrak{C}}(s) \coloneqq \norm{\varphi}_{\mathfrak{C}}(s)\ast^{\mathbb{A}}\norm{\psi}_{\mathfrak{C}}(s)$ for $\ast\in\{  \lor,\land,\&,\to\}$ and its corresponding algebraic operation $\ast^{\mathbb{A}}\in\{\lor,\land,\odot,\to\}$ on $\mathbb{A}$,
    \item $\norm{\varhexagon_{\lambda}\varphi}_{\mathfrak{C}}(s)  \coloneqq \lambda_S'(\norm{\varphi}_{\mathfrak{C}})(\sigma_V(s))$, for any predicate lifting $\lambda \in \Lambda$, and
    \item $\norm{\varhexagon_{\lambda^p}}_{\mathfrak{C}}(s)\coloneqq\lambda_S^p(!_S)(\sigma_V(s))=V(s)(p)$, for any $p\in P.$
\end{itemize}
\end{defn}
Given a set of formulas $\Gamma\subseteq\mathcal{ML}$, we use $\norm{\Gamma}_{\mathfrak{C}}(s)$ to denote the set $\{\norm{\varphi}_{\mathfrak{C}}(s)\;|\;\varphi\in\Gamma\}$ for any $s\in S$.

\begin{defn}
Let $\Gamma\cup\{\varphi\}\subseteq\mathcal{ML}$ and let $\mathfrak{C}=\langle S,\sigma_\nu\rangle$ be a $T$-model. Then, $\Gamma\Vdash_{\mathfrak{C}}\varphi$ denote that for all $s\in S$, $\norm{\Gamma}_{\mathfrak{C}}(s) =\{1\}$  implies $ \norm{\varphi}_{\mathfrak{C}}(s) = 1$. If $\Gamma\Vdash_{\mathfrak{C}}\varphi$ for all $T$-models $\mathfrak{C}$, then we say that $\varphi$ is a semantic consequence of $\Gamma$ and denote it by $\Gamma\Vdash\varphi$. When $\Gamma=\emptyset$, we simply abbreviate the notations as $\Vdash_{\mathfrak{C}}\varphi$ and $\Vdash\varphi$ respectively.
\end{defn}

Besides the general semantics for $\mathcal{ML}$, we can also define the step-$n$ semantic for the rank-$n$ language $\mathcal{L}_n$.
\begin{defn}
Let $\mathbf{1}=\{\bullet\}$ be a singleton set (i.e. a terminal object) in $\mathbf{Set}$. A {\em (pseudo)-terminal sequence\/} based on the functor $T_P$ is defined by induction:
\[\begin{array}{l}T_P^0\tilde{\mathbf{1}}=\tilde{\mathbf{1}}:=Hom(P,A)\times\mathbf{1},\\
T_P^n\tilde{\mathbf{1}}=T_P(T_P^{n-1}\tilde{\mathbf{1}})\;\;\mbox{\rm if}\;\;n>0.\end{array}\]
Then, for any $n\geq 0$ and  $\varphi\in\mathcal{L}_n$, its $n$-step semantics $\norm{\varphi}_n:T_P^n\tilde{\mathbf{1}}\to A$ is defined as follows: for all $\langle\nu,\delta\rangle\in T_P^n\tilde{\mathbf{1}}$
\begin{itemize}
    \item $\norm{\bar{c}}_n(\langle \nu,\delta\rangle) \coloneqq c$ with $c \in A$,
    \item $\norm{\varphi\ast \psi}_n(\langle\nu,\delta\rangle) \coloneqq \norm{\varphi}_n(\langle\nu,\delta\rangle)\ast^{\mathbb{A}}\norm{\psi}_n(\langle\nu,\delta \rangle)$ for $\ast\in\{\lor,\land, \&, \to\}$ and its corresponding algebraic operation $\ast^{\mathbb{A}}\in\{\lor,\land,\odot,\to\}$ $\mathbb{A}$,
    \item $\norm{\varhexagon_\lambda \varphi}_n(\langle \nu,\delta\rangle) \coloneqq \lambda_{T_P^{n-1}\tilde{\mathbf{1}}}' (\norm{\varphi}_{n-1})(\langle \nu,\delta\rangle)$, where $\lambda \in \Lambda$ is a predicate lifting.
    \item $\norm{\varhexagon_{\lambda^p}}_n(\langle\nu,\delta\rangle)\coloneqq \nu(p)$
\end{itemize}
\end{defn}
Note that our definition of terminal sequence is a little unusual because its starting point $T_P^0\tilde{\mathbf{1}}$ is not the terminal object $\mathbf{1}$  but $\tilde{\mathbf{1}}$. Such an unusuality is necessary to ensure the well-definedness of $\norm{\varhexagon_{\lambda^p}}_0$ for any propositional symbol $p\in{\mathcal L}_0$.  As above, for a  set of formulas $\Gamma\subseteq\mathcal{L}_n$, we use the notation $\norm{\Gamma}_n(t)$ to denote the set $\{\norm{\varphi}_n(t)\;|\;\varphi\in\Gamma\}$ for any $t\in T_P^n\tilde{\mathbf{1}}$.
\begin{defn}
Let $\Gamma\cup\{\varphi\}\subseteq\mathcal{L}_n$. We say that $\varphi$ is a step-$n$ semantic consequence of $\Gamma$, denoted by $\Gamma\Vdash_n\varphi$, if $\norm{\Gamma}_n(t)=\{1\}$ implies $\norm{\varphi}_n(t)=1$ for all $t\in T_P^n\tilde{\mathbf{1}}$. When $\Gamma=\emptyset$, we simply write $\Gamma\Vdash_n\varphi$ as $\Vdash_n\varphi$.
\end{defn}

Given a $T_P$-coalgebra $\langle S,\sigma_V \rangle$, let us define $\sigma_k$ for $k\in\omega$ inductively as follows
\[\begin{array}{l}\sigma_0:S\to\tilde{\mathbf{1}},\;\; s\mapsto\langle V(s),\bullet\rangle,\\
\sigma_k=T_P\sigma_{k-1}\circ\sigma_V.\end{array}\]
Then, we have the following connection between $\norm{\cdot}_{\mathfrak{C}}$ and $\norm{\cdot}_n$.
\begin{thm}\label{thm1}
For any $\varphi\in\mathcal{L}_n$, we have $\norm{\varphi}_{\mathfrak{C}}=\norm{\varphi}_n\circ\sigma_n$.
\end{thm}
\begin{proof}
We prove this theorem with simultaneous induction on $n$ and on the complexity of $\varphi \in \mathcal{L}_n$ for. We first fix a state $s \in S$.
\begin{enumerate}
\item $n=0$: by induction on the complexity of $\pi\in\mathcal{L}_0$
\begin{enumerate}
\item $\pi=\bar{c}$ for $c\in A$: $\norm{\bar{c}}_{\mathfrak{C}}(s)=c= \norm{\bar{c}}_0(\sigma_0(s))$
\item $\pi=\varhexagon_{\lambda^p}$ for a propositional symbol $p$: we have
\begin{align*}
    \norm{\varhexagon_{\lambda^p}}_{\mathfrak{C}}(s) & = \lambda_S^p(!_S)(\sigma_V(s))\\
    & = V(s)(p)\\
    & = \norm{\varhexagon_{\lambda^p}}_0(\langle V(s),\bullet\rangle)\\
    & = \norm{\varhexagon_{\lambda^p}}_0(\sigma_0(s))\\
\end{align*}
\item $\pi=\pi_0\ast\pi_1$ for $\pi_0,\pi_1\in\mathcal{L}_0$ and logical connectives $\ast$: by the inductive hypothesis, we have  \begin{align*}
    \norm{\pi_0\ast\pi_1}_{\mathfrak{C}}(s) & =\norm{\pi_0}_{\mathfrak{C}}(s)\ast^{\mathbb{A}}\norm{\pi_1}_{\mathfrak{C}}(s)\\
    & = \norm{\pi_0}_0(\sigma_0(s)) \ast^{\mathbb{A}} \norm{\pi_1}_0(\sigma_0(s))\\
    & = \norm{\pi_0 \ast \pi_1}_0(\sigma_0(s))\\
\end{align*}
\end{enumerate}
\item $n\geq 1$:  by induction on the complexity of $\varphi\in\mathcal{L}_n$
\begin{enumerate}
\item $\varphi=\bar{c}$ or $\varphi=\varphi_0\ast\varphi_1$: the proof is the same as in the case of $n=0$
\item $\varphi=\varhexagon_{\lambda^p}$: Recall that $\sigma_n(s)=(\text{id}_{Hom(P,A)}\times T\sigma_{n-1})(\sigma_V(s))=(V(s),T\sigma_{n-1}(\sigma(s)))$. Hence,
\begin{align*}
    \norm{\varhexagon_{\lambda^p}}_{\mathfrak{C}}(s) & = \lambda_S^p(!_S)(\sigma_V(s))\\
    & = V(s)(p)\\
    & = \norm{\varhexagon_{\lambda^p}}_n(\sigma_n(s))\\
\end{align*}
\item $\varphi=\varhexagon_{\lambda}\psi$: For simplicity, we abbreviate Hom-functors $Hom(-,A)$ and $Hom(T_P(-),A)$ as $H$ and $HT_P$ respectively in the following derivation. Recall that for the functor $H$, we have $Hf(g)=g\circ f$ for morphisms $f$ and $g$ of appropriate types. Then, we have
\[\begin{array}{ll}
    \norm{\varhexagon_{\lambda}\psi}_{\mathfrak{C}}(s)=\lambda_S'(\norm{\psi}_{\mathfrak{C}})(\sigma_V(s))&\mbox{\rm Def. 5}\\
    =\lambda_S'(\norm{\psi}_{n-1}\circ\sigma_{n-1})(\sigma_V(s))&\mbox{\rm ind. hyp.}\\
    =\lambda_S'(H\sigma_{n-1}(\norm{\psi}_{n-1}))(\sigma_V(s))&g\circ f=Hf(g)\\
    =[\lambda_S' \circ H\sigma_{n-1}](\norm{\psi}_{n-1})(\sigma_V (s))&\mbox{\rm rewritting} \\
    =[HT_P\sigma_{n-1}\circ\lambda_{T^{n-1}_P\tilde{\mathbf{1}}}'](\norm{\psi}_{n-1})(\sigma_V(s))& \mbox{\rm naturality of}\;\; \lambda'\\
    =[H(T_P\sigma_{n-1})(\lambda_{T^{n-1}_P\tilde{\mathbf{1}}}'(\norm{\psi}_{n-1}))](\sigma_V(s))& \mbox{\rm rewritting} \\
    =[\lambda_{T^{n-1}_P\tilde{\mathbf{1}}}'(\norm{\psi}_{n-1})\circ(T_P\sigma_{n-1}\circ\sigma_V)](s)& Hf(g)=g\circ f\\
    =[\lambda_{T^{n-1}_P\tilde{\mathbf{1}}}'(\norm{\psi}_{n-1})\circ\sigma_n](s)& \mbox{\rm def. of}\;\; \sigma_n\\
    =[\norm{\varhexagon_{\lambda}\psi}_n \circ \sigma_n ](s)& \mbox{\rm Def. 7}
\end{array}
\]
\end{enumerate}
\end{enumerate}
\end{proof}

Recall that there is a unique map $!_X:X\to\mathbf{1}$ for any set $X$ because $\mathbf{1}$ is the terminal object. Hence, if $Y$ is a set of the form $Hom(P,A)\times X$, we can define a surjective map $!'_Y:Y\to\tilde{\mathbf{1}}$ as $!'_Y:=\text{id}_{Hom(P,A)}\times !_X$. In particular, when $Y=T_P\tilde{\mathbf{1}}$, we write the surjection $!'_{T_P\tilde{\mathbf{1}}}$ as $\gamma^0$ and it has a right inverse $\iota^0:\tilde{\mathbf{1}}\to T_P\tilde{\mathbf{1}}$ so that $\gamma^0 \circ \iota^0 = \text{id}_{\tilde{\mathbf{1}}}$. For $n\geq 1$, let $\gamma^n=T_P^n\gamma^0 : T_P^{n+1}\tilde{\mathbf{1}}\to T_P^n\tilde{\mathbf{1}}$, and $\iota^n = T_P^n\iota^0:T_P^n\tilde{\mathbf{1}}\to T_P^{n+1}\tilde{\mathbf{1}}$. We obtain $\gamma^n \circ \iota^n = \text{id}_{T_P^n\tilde{\mathbf{1}}}$. Note that $\langle T_P^n\tilde{\mathbf{1}}, \iota^n\rangle$ is a $T_P$-coalgebra for every $n\in\omega$.  Hence, by respectively substituting $S$ and $\sigma_V$ with $T_P^n\tilde{\mathbf{1}}$ and $\iota^n$ in the definition of the sequence $(\sigma_k)_{k\in\omega}$ above,  we can also define $\iota^n_k$ for $k\in\omega$ inductively
$$\begin{array}{l}\iota_0^n:T_P^n\tilde{\mathbf{1}}\to \tilde{\mathbf{1}}; (\nu,\delta)\mapsto(\nu,\bullet)\\
\iota_k^n = T_P\iota^n_{k-1}\circ\iota^n.\end{array}$$.

\begin{lem}\label{lem1}
For all $k\leq n$, $\iota^n_k=T_P^k(!'_{T_P^{n-k}\tilde{\mathbf{1}}})$.
\end{lem}
\begin{proof}
The proof is similar to that for Lemma 4.11 in \cite{pattinson2003coalgebraic} and we omit it here.
\end{proof}
\begin{thm}\label{thm3}
Suppose  $\varphi \in \mathcal{L}_n$, we have $\Vdash\varphi$ iff $\Vdash_n\varphi$
\end{thm}
\begin{proof}
For the only if direction, let us consider the coalgebra $\mathfrak{C}=\langle T_P^n\tilde{\mathbf{1}},\iota^n\rangle$. Then, by the assumption, $\norm{\varphi}_{\mathfrak{C}}(t)=1$ for all $t\in T_P^n\tilde{\mathbf{1}}$. By Lemma\ref{lem1}, we have $\iota^n_n=\text{id}_{T_P^n\tilde{\mathbf{1}}}$. Hence, by Theorem\ref{thm1}, we have $\norm{\varphi}_n(t)=\norm{\varphi}_n\circ\iota^n_n(t)=\norm{\varphi}_{\mathfrak{C}}(t)=1$ for all $t\in T_P^n\tilde{\mathbf{1}}$.

For the other direction, suppose that $\Vdash_n\varphi$. Then, for all $t \in T_P^n\tilde{\mathbf{1}}$, we have $\norm{\varphi}_n(t)= 1$. Hence, by Theorem \ref{thm1}, for any $T_P$-coalgebra $\mathfrak{C}=\langle S,\sigma_V\rangle$, we can construct the sequence $(\sigma_k)_{k\in\omega}$ such that for any $s \in S$, $\norm{\varphi}_{\mathfrak{C}}(s)=\norm{\varphi}_n(\sigma_n(s))=1$.
\end{proof}

\begin{cor}\label{thm2}
Suppose  $\Gamma\cup\{\varphi\}\in\mathcal{L}_n$, we have $\Gamma\Vdash\varphi$ iff $\Gamma\Vdash_n\varphi$
\end{cor}
\begin{proof}
The proof is a straightforward extension of that for Theorem \ref{thm3}.
\end{proof}

\section{Proof system}\label{sec4}
In this section, we characterize semantic consequences in $\mathcal{ML}$ and its rank-$n$ fragment with provability relations. First, we  introduce the concept of basic derivation relation, called a consecution.
\begin{defn}
A {\it consecution} in $\mathcal{ML}$ is a pair $\langle\Gamma,\varphi\rangle$, where $\Gamma$ is a finite set of formulas of $\mathcal{ML}$ and $\varphi\in\mathcal{ML}$. We usually write a consecution $\langle\Gamma,\varphi\rangle$ as $\Gamma\vdash\varphi$ and for a set of consecutions $\mathbf{L}$, we always abbreviate $\Gamma\vdash\varphi\in\mathbf{L}$ as $\Gamma\vdash_{\mathbf{L}}\varphi$ and omit $\Gamma$ when it is empty.
\end{defn}
For a set of consecutions to be a nice logical system, we need to impose some closure properties on it. As usual, an axiom is regarded as a schema. Hence, a logical system should contain all substitution instances of its axioms. More precisely, a {\it substitution} is a map $\rho: P \to\mathcal{ML}$. We will use the notation $(\varphi_i/p_i: p_i\in I)$ for the substitution that maps each variable $p_i\in I$ to the formula $\varphi_i$ and remain identical in other variables $p\in P\backslash I$, where $I$ is a (typically finite) subset of $P$. The application of a substitution $\rho=(\varphi_i/p_i: p_i\in I)$ to a formula $\varphi$ results in a new formula $\varphi\rho$ in which the nullary modality $\varhexagon_{\lambda^{p_{i}}}$ is uniformly replaced by $\varphi_i$ for each $p_i\in I$ and we say that $\varphi\rho$ is an instance of $\varphi$.  An instance of a consecution $\langle\Gamma,\varphi \rangle$ is $\langle\Gamma\rho,\varphi\rho\rangle$ where $\Gamma\rho:=\{\varphi\rho\;|\;\varphi\in\Gamma\}$. If the range of a substitution $\rho$ is confined to $\mathcal{L}_n$, i.e.\ $\rho:P\to\mathcal{L}_n$, then we call $\rho$ an {\it n-substitution}.

As we have assumed the existence of a logical axiomatization $Ax(\mathbb{A})$ for the FL algebra $\mathbb{A}$, we can regard  the derivability of $\varphi$ from $\Gamma$ in $Ax(\mathbb{A})$ as a consecution and denote it by $\Gamma\vdash_{Ax(\mathbb{A})}\varphi$. Let us also fix a set of consecutions $Ax(\Lambda)\subseteq\mathbb{P}(\mathcal{L}_1)\times\mathcal{L}_1$ for rank-1 modal axioms. Then, we can define the general consequence relation as follows.
\begin{defn}
A set of consecutions $\mathbf{L}$ is called a {\it general consequence relation} if it is the least set satisfying the following closure properties
\begin{itemize}
    \item If $\Gamma\vdash_{Ax(\mathbb{A})}\varphi$, then $\Gamma\vdash_{\mathbf{L}}\varphi$.
    \item If $\Gamma\vdash_{Ax(\Lambda)}\varphi$, then $\Gamma\rho\vdash_{\mathbf{L}}\varphi\rho$ for each substitution $\rho$.
    \item If $\Gamma\vdash_{\mathbf{L}}\varphi$, then $\varhexagon_{\lambda}\Gamma\vdash_{\mathbf{L}}\varhexagon_{\lambda}\varphi$ for each  $\lambda\in\Lambda$ where $\varhexagon_{\lambda}\Gamma:=\{\varhexagon_{\lambda}\varphi\ \;|\; \varphi\in\Gamma\}$.
\end{itemize}
\end{defn}
The role of modal axioms in $Ax(\Lambda)$ is to characterize predicate liftings in $\Lambda$. In the general framework, we do not restrict to any specific functor $T$ and predicate liftings. Thus, we do not consider a concrete set of axioms in $Ax(\Lambda)$. For instance, one might include $\langle\emptyset,\Box\bar{1}\rangle$ or $\langle\{\Box\varphi\wedge\Box\psi\},\Box(\varphi\wedge\psi)\rangle$ in $Ax(\Lambda)$ for a particular predicate lifting $\Box$. In addition, because $\mathbf{L}$ includes the derivation relation in $Ax(\mathbb{A})$, it must include all instances of axioms in $Ax(\mathbb{A})$ and be closed under inference rules of $Ax(\mathbb{A})$. For example, $\mathbf{L}$ should be closed under the following rule if Modus Ponens is an inference rule of $Ax(\mathbb{A})$.
\begin{center}
    \AxiomC{$\vdash_{\mathbf{L}}\varphi$}
    \AxiomC{$\vdash_{\mathbf{L}}\varphi\to\psi$}
\RightLabel{(MP)}
    \BinaryInfC{$\vdash_{\mathbf{L}} \psi$}
\DisplayProof
\end{center}

Since $\mathcal{ML}$ can be stratified into the union of $\mathcal{L}_n$ for all $n \in \omega$, we can also define the step-$n$ consequence relation $\mathbf{L}_n$ for each $\mathcal{L}_n$.
\begin{defn}\badstart
\begin{enumerate}
\item $n=0$:We define the step-0 consequence relation $\mathbf{L}_0\subseteq\mathbb{P}(\mathcal{L}_0)\times\mathcal{L}_0$ by $\Gamma\vdash_{\mathbf{L}_0}\varphi$ iff $\Gamma\vdash_{Ax(\mathbb{A})}\varphi$ for any $\Gamma\cup\{\varphi\}\subseteq\mathcal{L}_0$.
\item $n>0$: A set of consecutions $\mathbf{L}_n\subseteq\mathbb{P}(\mathcal{L}_n)\times \mathcal{L}_n$ forms the {\it step-$n$ consequence relation} if it is the least set satisfying the following closure properties:
    \begin{itemize}
    \item If $\Gamma\vdash_{Ax(\mathbb{A})}\varphi$, then $\Gamma\vdash_{\mathbf{L}_n}\varphi$ for any $\Gamma\cup\{\varphi\}\subseteq\mathcal{L}_n$.
    \item If $\Gamma\vdash_{Ax(\Lambda)}\varphi$, then $\Gamma\rho\vdash_{\mathbf{L}_n}\varphi\rho$ for any $(n-1)$-substitution $\rho$ and $\Gamma\cup\{\varphi\}\subseteq\mathcal{L}_1$.
    \item If $\Gamma\vdash_{\mathbf{L}_{n-1}}\varphi$, then $\varhexagon_{\lambda}\Gamma\vdash_{\mathbf{L}_n}\varhexagon_{\lambda}\varphi$ for any $\lambda\in\Lambda$ and $\Gamma\cup\{\varphi\}\subseteq\mathcal{L}_{n-1}$.
\end{itemize}
\end{enumerate}
\end{defn}

As there is an intimate connection between $\Vdash$ and $\Vdash_n$, we can also find an analogous relationship between general and step-$n$ consequence relations.
\begin{thm}\label{thm4}
Suppose $\Gamma\subseteq\mathcal{L}_n$ and $\varphi\in\mathcal{L}_n$. Then $\Gamma\vdash_{\mathbf{L}}\varphi$ iff $\Gamma\vdash_{\mathbf{L}_n}\varphi$
\end{thm}
\begin{proof}
It is obvious that $\mathbf{L}_n\subseteq\mathbf{L}$ for any $n\in\omega$. Thus, we only need to prove that $\Gamma\vdash_{\mathbf{L}}\varphi$ implies  $\Gamma\vdash_{\mathbf{L}_n}\varphi$. We prove this by induction on n. For $n=0$, the only way that $\Gamma\vdash_{\mathbf{L}}\varphi$ is $\Gamma\vdash_{Ax(\mathbb{A})}\varphi$. Therefore, $\Gamma\vdash_{\mathbf{L}_0} \varphi$ follows by definition. For $n>0$, let us consider all possible ways for the derivation of $\Gamma\vdash_{\mathbf{L}}\varphi$. First, if $\Gamma\vdash_{Ax(\mathbb{A})}\varphi$, then $\Gamma\vdash_{\mathbf{L}_n}\varphi$ holds by definition. Second, if $\Gamma\vdash_{\mathbf{L}}\varphi$ is a substitution instance of $Ax(\Lambda)$, then the substitution must be an $(n-1)$-substitution because of the form of modal axioms. Hence, we have $\Gamma\vdash_{\mathbf{L}_n}\varphi$. Finally, if $\Gamma \vdash_{\mathbf{L}}\varphi$ was derived by $\Gamma'\vdash_{\mathbf{L}}\psi$ such that  $\Gamma=\varhexagon_{\lambda}\Gamma'$ and $\varphi=\varhexagon_{\lambda}\psi$ for some $\Gamma'\cup\{\psi\}\subseteq\mathcal{L}_{n-1}$, then by the  inductive hypothesis, we have $\Gamma' \vdash_{\mathbf{L}_{n-1}}\psi$, which leads to the desired result.
\end{proof}

\section{One-step Logic: Soundness and Completeness}\label{sec5}
We have defined both logical systems and semantics for ${\mathcal ML}$ and its rank-$n$ fragments. Now, we can consider the soundness and completeness of a logical system with respect to its corresponding semantics. As usual, we say that $\vdash$ is (finitely) sound (resp.\ complete) with respect to $\Vdash$ if $\Gamma\vdash\varphi$ implies $\Gamma\Vdash\varphi$(resp.\ $\Gamma\Vdash\varphi$ implies $\Gamma\vdash\varphi$) for any finite subsets of formulas $\Gamma\cup\{\varphi\}$. By the definition of step-$n$  consequence relation, we know that $\mathbf{L}_0$ is simply the derivation relation of the underlying many-valued logic, which is sound and complete with respect to $\Vdash_0$\cite{cintula2013two}. For $n>0$, we define the one-step version of soundness and completeness for $\mathcal{L}_n$.
\begin{defn}
We say that $\vdash_{\mathbf{L}_n}$ is {\it  one-step sound (resp.\ complete)} if
the soundness (resp.\ completeness) of $\vdash_{\mathbf{L}_{n-1}}$ with respect to $\Vdash_{n-1}$ implies that of $\vdash_{\mathbf{L}_n}$ with respect to $\Vdash_n$. In addition, the general consequence relation $\vdash_{\mathbf{L}}$ is one-step sound (resp.\ complete) if $\vdash_{\mathbf{L}_n}$ is one-step sound (resp.\ complete) for each $n>0$.
\end{defn}

Because of the full generality of the definition, to check the one-step soundness and completeness of a system will depend on the specification of the particular functor and predicate liftings, as well as their characteristic modal axioms. 

Although we do not provide any concrete logical system here, we indeed have some general conditions on  $Ax(\Lambda)$ and $\Lambda$ that can sufficiently guarantee the one-step soundness of a system as in the classical case~\cite{pattinson2003coalgebraic}. First, we require that $Ax(\Lambda)$ is sound in a step-wise way.
\begin{defn}
The set of modal axioms $Ax(\Lambda)$ is {\it step-$n$ sound} if for any $ \Gamma\vdash_{Ax(\Lambda)}\varphi$ and $(n-1)$-substitution $\rho$, we have $\Gamma\rho\Vdash_n\varphi\rho$.
\end{defn}
Second, we need a condition on predicate liftings corresponding to order-preserving in the classical case~\cite{pattinson2003coalgebraic}. To do that, we note that the denotation of many-valued logic formulas can be regarded as (the membership function of) an $\mathbb{A}$-valued fuzzy subset. That is, it is a function from an universe $X$ to $A$. Then, for any such a function $f:X\to A$ and $\alpha\in A$, we can define the {\em $\alpha$-cut\/} of $f$ as $f_\alpha=\{x\in X\mid f(x)\geq^\mathbb{A}\alpha\}$. In addition, for any two  families of $\mathbb{A}$-valued fuzzy subsets $F,G$ over the same universe, we define the ordering $F\sqsubseteq_\alpha G$ as $\bigcap_{f\in F}f_\alpha\subseteq \bigcup_{g\in G}g_\alpha$. 
\begin{defn}
Let $S$ be any set and $\alpha\in A$. A predicate lifting $\lambda: Hom(-,A)\Rightarrow Hom(T(-),A)$ is called {\it $\alpha$-preserving} if for any two families of $\mathbb{A}$-valued fuzzy subsets $F,G$ over $S$ such that, $F\sqsubseteq_\alpha G$, we have $\lambda_SF\sqsubseteq_\alpha\lambda_SG$, where $\lambda_SF=\{\lambda_S(f)\mid f\in F\}$ and $\lambda_SG$ is defined in the same way.
\end{defn}

With these definitions, we can provide the following sufficient condition for the one-step soundness of $\vdash_{\mathbf{L}_n}$.
\begin{prop}\label{prop1}
Suppose that $Ax(\Lambda)$ is step-$n$ sound and every $\lambda\in\Lambda$ is 1-preserving. Then, $\vdash_{\mathbf{L}_n}$ is one-step sound.
\end{prop}
\begin{proof}
Assume the soundness of $\vdash_{\mathbf{L}_{n-1}}$, we can prove that of $\vdash_{\mathbf{L}_n}$ by enumerating all possible ways of derivations in it. First, by the soundness of $\mathfrak{L}(A)$, the derivation of $\Gamma\vdash_{\mathbf{L}_n}\varphi$ from axioms and rules of  $Ax(\mathbb{A})$ implies $\Gamma\Vdash_n\varphi$. Second, if $\Gamma\vdash_{\mathbf{L}_n} \varphi$ is an instance of $Ax(\Lambda)$, $\Gamma\Vdash_n\varphi$ holds by the step-$n$ soundness of $Ax(\Lambda)$. Finally, if $\Gamma\vdash_{\mathbf{L}_n}\varphi$ was derived by adding modality, where $\Gamma=\varhexagon_{\lambda}\Gamma'$ and $\varphi=\varhexagon_{\lambda}\psi$ for $\Gamma'\cup\{\psi\}\subseteq \mathcal{L}_{n-1}$ such that $\Gamma'\vdash_{\mathbf{L}_{n-1}}\psi$. then, because $\vdash_{\mathbf{L}_{n-1}}$ is sound with respect to $\Vdash_{n-1}$, we have $\Gamma'\Vdash_{n-1}\psi$, which implies $\Gamma\Vdash_n\varphi$ by the 1-preservation of $\lambda$.
\end{proof}

As for the one-step completeness, its sufficient conditions in the classical case depend on the admissibility and  reflexivity of $Ax(\Lambda)$  \cite{pattinson2003coalgebraic}. However, the applicability of these conditions requires the transformation of propositional formulas into conjunctive or disjunctive normal forms which is a property that most of many-valued logics do not enjoy. Hence, it is still an open issue to find suitable conditions for guaranteeing  the one-step completeness of a many-valued coalgebraic modal logic.

Next, we show that the soundness and completeness of many-valued coalgebraic modal logics can be determined at the one-step level.
\begin{thm}\label{thm7}
If $\vdash_\mathbf{L}$ is one-step sound (resp.\ complete), then  $\vdash_{\mathbf{L}_n}$ is sound (resp.\ complete) with respect to $\Vdash_n$ for each $n\in\omega$.
\end{thm}
\begin{proof}
The proof is an easy induction on n. For $n=0$, the claim follows from the soundness and completeness of $Ax(\mathbb{A})$. For $n>0$,  one-step soundness and completeness of $\vdash_\mathbf{L}$ implies that the soundness and completeness of $\vdash_{\mathbf{L}_{n-1}}$ can be transferred to the next level inductively.
\end{proof}

\begin{cor}
If $\vdash_\mathbf{L}$ is one-step sound (resp.\ complete), then  $\vdash_{\mathbf{L}}$ is  sound (resp.\ complete) with respect to $\Vdash$.
\end{cor}
\begin{proof}
Because $\mathcal{ML}=\bigcup_{n\in\omega}\mathcal{L}_n$, for any finite set $\Gamma\cup\{\varphi\}\subseteq\mathcal{ML}$, we may assume that $\Gamma\cup\{\varphi\}\subseteq\mathcal{L}_n$ for some $n$. Hence, we have
\[\Gamma\vdash_\mathbf{L}\varphi\Leftrightarrow\Gamma\vdash_{\mathbf{L}_n}\varphi\Leftrightarrow\Gamma\Vdash_n\varphi
\Leftrightarrow\Gamma\Vdash\varphi\]
by Theorem~\ref{thm4}, Theorem~\ref{thm7}, and Corollary~\ref{thm2}
\end{proof}

\section{Finite model property}\label{sec6}
In the previous section, we have seen the connection between coalgebraic model $\mathfrak{C}$ and the terminal sequence $(T_P^n\tilde{\mathbf{1}})_{n \in \omega}$. In this section, we will use this connection to prove the finite model property by imposing some additional assumption on the functor $T$. Recall that the finite model property states that if $\varphi\in\mathcal{ML}$ is satisfiable, then $\varphi$ is satisfiable in a finite $T$-model.
\begin{defn}
A $T_P$ functor is called {\it finite} if $TS$ is finite whenever $S$ is a finite set.
\end{defn}
A formula $\varphi\in\mathcal{ML}$ is {\it satisfiable} if there exists a $T$-model $\mathfrak{C}$ such that $\norm{\varphi}_{\mathfrak{C}}(s)=1$ for some $s\in S$.
\begin{lem}\label{lma2}
For any $\varphi\in\mathcal{L}_n$, if  $\varphi$ is satisfiable, then it is satisfiable in the $T$-model $\langle T_P^n\tilde{\mathbf{1}}, \iota^n\rangle$.
\end{lem}
\begin{proof}
According to the assumption, there exists a $T$-model $\mathfrak{C}=\langle S,\sigma_V\rangle$ with $\norm{\varphi}_{\mathfrak{C}}(s)=1$ for some $s\in S$. Since $\norm{\varphi}_{\mathfrak{C}}(s) = \norm{\varphi}_n(\sigma_n(s))=1$ by Theorem~\ref{thm1}, we have $\norm{\varphi}_{T_P^n\tilde{\mathbf{1}}}(\sigma_n(s))= \norm{\varphi}_n\circ\iota^n_n(\sigma_n(s)) =\norm{\varphi}_n(\sigma_n(s))=1 $. This proves the claim.
\end{proof}
\begin{cor}
Let $T_P$ be a finite functor defined above. Then, for each formula $\varphi\in\mathcal{ML}$, if $\varphi$ is satisfiable, then $\varphi$ is satisfiable in a finite $T$-model.
\end{cor}
\begin{proof}
Because $\mathcal{ML}=\bigcup_{n\in\omega}\mathcal{L}_n$, we might assume that $\varphi\in\mathcal{L}_n$ for some $n\in\omega$. By Lemma~\ref{lma2}, $\varphi$ is satisfiable in the $T$-model $\langle T_P^n\tilde{\mathbf{1}}, \iota^n\rangle$. The finiteness of $T_P^n\tilde{\mathbf{1}}$ follows from the finiteness of $T_P$ and $\tilde{\mathbf{1}}$, where $\tilde{\mathbf{1}}$ is finite because $Hom(P,A)$ is finite by the finiteness of  $T_P$.
\end{proof}

\section{Conclusion}\label{sec7}
We have used the stratification method and induction on the modal rank of formulas to prove that the soundness and completeness of many-valued coalgebraic modal logics are determined at the one-step level. Besides, we also prove finite model property under the finiteness assumption of the functor. This is different from  methods of canonical model construction and filtration employed in \cite{Lin2022}. As a consequence, we no longer require that the underlying many-valued language must have the expressive power to internalize the meta-level truth valuation operations, as it is needed during the construction of canonical model in \cite{Lin2022}.

In \cite{pattinson2010cut,venema2012completeness}, the stratification method is also used to prove either cut-free completeness for coalgebraic modal logics via predicate lifting or completeness for coalgebraic modal logics via relation lifting. To extend these results to the many-valued case, we need to consider an endofunctor $T$ over the slice category $\mathbf{Set}/Hom(P,A)$ rather than the product functor $T_P=EV_P\times T$ over the $\mathbf{Set}$ category. Furthermore, in this paper, we do not have a concrete example for many-valued modal logics which satisfy the assumptions for one-step soundness and completeness. Hence, to instantiate the framework to some many-valued modal logics is another pressing issue for further research.


\bibliographystyle{plain}
\bibliography{bibliography.bib}

\begin{thebibliography}{10}

\bibitem{Awodey}
S.~Awodey.
\newblock {\em Category Theory}.
\newblock Oxford University Press, 2nd edition, 2010.

\bibitem{bilkova2013many}
M.~B{\'\i}lkov{\'a} and M.~Dost{\'a}l.
\newblock Many-valued relation lifting and moss??coalgebraic logic.
\newblock In {\em International Conference on Algebra and Coalgebra in Computer
  Science}, pages 66--79. Springer, 2013.

\bibitem{bilkova2016expressivity}
M.~B{\'\i}lkov{\'a} and M.~Dost{\'a}l.
\newblock Expressivity of many-valued modal logics, coalgebraically.
\newblock In {\em International Workshop on Logic, Language, Information, and
  Computation}, pages 109--124. Springer, 2016.

\bibitem{hml}
P.~Blackburn, J.~van Benthem, and F.~Wolter.
\newblock {\em Handbook of Modal Logic}.
\newblock Elsevier, 2006.

\bibitem{bou2011minimum}
F.~Bou, F.~Esteva, L.~Godo, and R.O. Rodr{\'\i}guez.
\newblock On the minimum many-valued modal logic over a finite residuated
  lattice.
\newblock {\em Journal of Logic and computation}, 21(5):739--790, 2011.

\bibitem{fml13}
X.~Caicedo, G.~Metcalfe, R.O. Rodr\'{\i}guez, and J.~Rogger.
\newblock A finite model property for {G}{\"o}del modal logics.
\newblock In L.~Libkin, U.~Kohlenbach, and R.J.G.B. de~Queiroz, editors, {\em
  Proceedings of the 20th International Workshop on Logic, Language,
  Information, and Computation (WoLLIC)}, LNCS 8071, pages 226--237.
  Springer-Verlag, 2013.

\bibitem{fml10}
X.~Caicedo and R.O. Rodr\'{\i}guez.
\newblock Standard {G}{\"o}del modal logics.
\newblock {\em Studia Logica}, 94(2):189--214, 2010.

\bibitem{fml12}
X.~Caicedo and R.O. Rodr\'{\i}guez.
\newblock Bi-modal {G}{\"o}del logic over [0,1]-valued kripke frames.
\newblock {\em Journal of Logic and Computation}, 2012.

\bibitem{che}
B.F. Chellas.
\newblock {\em Modal Logic : An Introduction}.
\newblock Cambridge University Press, 1980.

\bibitem{cintula2011handbook}
P.~Cintula, P.~H{\'a}jek, and C.~Noguera.
\newblock Handbook of mathematical fuzzy logic (in 2 volumes), 2011.

\bibitem{cintula2013two}
P.~Cintula and C.~Noguera.
\newblock A general framework for mathematical fuzzy logic.
\newblock In P.~Cintula, P.~H{\'a}jek, and C.~Noguera, editors, {\em Handbook
  of Mathematical Fuzzy Logic-Volume 1 Studies in Logic, Mathematical Logic and
  Foundations}, pages 103--207. College Publications, 2011.

\bibitem{Cintula18}
P.~Cintula and C.~Noguera.
\newblock Neighborhood semantics for modal many-valued logics.
\newblock {\em Fuzzy Sets and Systems}, 345:99--112, 2018.

\bibitem{fitting1991many}
M.~Fitting.
\newblock Many-valued modal logics.
\newblock {\em Fundam. Inform.}, 15(3-4):235--254, 1991.

\bibitem{fitting1992many}
M.~Fitting.
\newblock Many-valued model logics ii.
\newblock {\em Fundam. Inform.}, 17(1-2):55--73, 1992.

\bibitem{Hajek07}
P.~H{\'{a}}jek.
\newblock Complexity of fuzzy probability logics {II}.
\newblock {\em Fuzzy Sets and Systems}, 158(23):2605--2611, 2007.

\bibitem{hajekGE95}
P.~H{\'{a}}jek, L.~Godo, and F.~Esteva.
\newblock Fuzzy logic and probability.
\newblock In P.~Besnard and S.~Hanks, editors, {\em Proceedings of the Eleventh
  Annual Conference on Uncertainty in Artificial Intelligence ({UAI}'95)},
  pages 237--244. Morgan Kaufmann, 1995.

\bibitem{kupke2015weak}
C.~Kupke and H.H. Hansen.
\newblock Weak completeness of coalgebraic dynamic logics.
\newblock In {\em FICS 2015: Proceedings of the 10th International Workshop on
  Fixed Points in Computer Science, Berlin, Germany, 11-12 September 2015}.
  Cornell university Library, 2015.

\bibitem{kupke2011coalgebraic}
C.~Kupke and D.~Pattinson.
\newblock Coalgebraic semantics of modal logics: an overview.
\newblock {\em Theoretical Computer Science}, 412(38):5070--5094, 2011.

\bibitem{Lin2022}
C.Y. Lin and C.J. Liau.
\newblock Many-valued coalgebraic modal logic: One-step completeness and finite
  model property, 2020.

\bibitem{lmcs11}
G.~Metcalfe and N.~Olivetti.
\newblock Towards a proof theory of {G}{\"o}del modal logics.
\newblock {\em Logical Methods in Computer Science}, 7(2):1--27, 2011.

\bibitem{moss1999coalgebraic}
L.S. Moss.
\newblock Coalgebraic logic.
\newblock {\em Annals of Pure and Applied Logic}, 96(1-3):277--317, 1999.

\bibitem{Ono2003}
H.~Ono.
\newblock Substructural logics and residuated lattices --- an introduction.
\newblock In V.F. Hendricks and J.~Malinowski, editors, {\em Trends in Logic:
  50 Years of Studia Logica}, pages 193--228. Springer Netherlands, 2003.

\bibitem{Ostermann88}
P.~Ostermann.
\newblock Many-valued modal propositional calculi.
\newblock {\em Mathematical Logic Quarterly}, 34(4):343--354, 1988.

\bibitem{Ostermann90}
P.~Ostermann.
\newblock Many-valued modal logics: Uses and predicate calculus.
\newblock {\em Mathematical Logic Quarterly}, 36(4):367--376, 1990.

\bibitem{pattinson2003coalgebraic}
D.~Pattinson.
\newblock Coalgebraic modal logic: Soundness, completeness and decidability of
  local consequence.
\newblock {\em Theoretical Computer Science}, 309(1-3):177--193, 2003.

\bibitem{pattinson2010cut}
D.~Pattinson and L.~Schr{\"o}der.
\newblock Cut elimination in coalgebraic logics.
\newblock {\em Information and Computation}, 208(12):1447--1468, 2010.

\bibitem{schroder2007finite}
L.~Schr{\"o}der.
\newblock A finite model construction for coalgebraic modal logic.
\newblock {\em The Journal of Logic and Algebraic Programming},
  73(1-2):97--110, 2007.

\bibitem{SchroderP09}
L.~Schr{\"{o}}der and D.~Pattinson.
\newblock Strong completeness of coalgebraic modal logics.
\newblock In S.~Albers and J.Y. Marion, editors, {\em Proceedings of the 26th
  International Symposium on Theoretical Aspects of Computer Science}, volume~3
  of {\em LIPIcs}, pages 673--684. Schloss Dagstuhl - Leibniz-Zentrum f{\"{u}}r
  Informatik, Germany, 2009.

\bibitem{SchroderP11}
L.~Schr{\"{o}}der and D.~Pattinson.
\newblock Description logics and fuzzy probability.
\newblock In {\em Proceedings of the 22nd International Joint Conference on
  Artificial Intelligence ({IJCAI})}, pages 1075--1081. AAAI, 2011.

\bibitem{venema2012completeness}
Y.~Venema, A.~Kurz, and C.~Kupke.
\newblock Completeness for the coalgebraic cover modality.
\newblock {\em Logical Methods in Computer Science}, 8, 2012.

\end{thebibliography}

\newpage
\appendix
\section{Basic Notions of Category Theory}
In this appendix, we review some basic notions of category by following the presentation in \cite{Awodey}.
\begin{defn}
A category $\mathsf{C}$ consists of the following components
\begin{itemize}
	\item A collection of {\em objects}: $Ob(\mathsf{C})=\{A, B, C, \cdots\}$
	\item A collection of {\em arrows (morphisms)}: $Ar(\mathsf{C})=\{f, g, h, \cdots\}$
	\item For each arrow $f$, there are given objects $A=dom(f)$ and $B=cod(f)$, called the domain and codomain of $f$ respectively. We use $f:A\to B$ to indicate an arrow with its {\em domain} and {\em codomain} at the same time.
	\item For arrows $f:A\to B$ and $g:B\to C$, there exists an arrow $g\circ f:A\to C$, called the {\em composite} of $f$ and $g$.
	\item For each object $A$, there is an arrow $1_A:A\to A$, called the {\em identity arrow} of $A$.
	\item Identity arrows and composites are required to satisfy the following laws:
	\begin{itemize}
		\item Associativity: for all $f:A\to B, g:B\to C$, and $h:C\to D$,\[h\circ (g\circ f)=(h\circ g)\circ f\]
		\item Unit: for all $f:A\to B$\[f\circ 1_A=f=1_B\circ f.\]
	\end{itemize}	
\end{itemize}
\end{defn}
We use $\mathsf{C}, \mathsf{D}$, etc.\ to denote a category. A particular example of category used in this paper is $\mathsf{Set}$, whose objects and morphisms are simply sets and functions respectively. The composite of two functions is their functional composition and the identity arrow of a set is the identity function on it. The {\em opposite\/} category $\mathsf{C}^{op}$ of a category $\mathsf{C}$ has the same objects as $\mathsf{C}$ and an arrow $f:C\to D$ in $\mathsf{C}^{op}$ is an arrow $f:D\to C$ in $\mathsf{C}$.

Given two objects $A$ and $B$ in any category $\mathsf{C}$, we write $Hom(A,B)=\{f\in Ar(\mathsf{C})\mid f:A\to B\}$ and call such a set of arrows as a Hom-set. Note that Hom-set is an object of the category $\mathsf{Set}$ by definition.

Just like functions play the role of arrows between sets, there is a corresponding notion of mappings between categories.
\begin{defn}
A functor $F:\mathsf{C}\to\mathsf{D}$ between two  categories $\mathsf{C}$ and $\mathsf{D}$ is defined as a mapping of objects to objects and arrows to arrows such that
\begin{itemize}
	\item if $f:A\to B$ is an arrow in $Ar(\mathsf{C})$, then $F(f):F(A)\to F(B)$ is an arrow in $Ar(\mathsf{D})$,
	\item for arrows $f:A\to B$ and $g:B\to C$ in $Ar(\mathsf{C})$, $F(g\circ f)=F(g)\circ F(f)$, and
	\item for each object $A\in Ob(\mathsf{C})$, $F(1_A)=1_{F(A)}$
\end{itemize}
\end{defn}
We sometimes omit the parentheses in the application of a functor to objects or arrows in case of no ambiguity. That is, we will usually write $FA$ or $Ff$ instead of $F(A)$ or $F(f)$. In addition, for the successive application of functors, we can write $FGA$ and $FGf$ to denote $F(G(A))$ and $F(G(f))$ respectively.

Because the opposite category $\mathsf{C}^{op}$  has the same objects as $\mathsf{C}$, a functor from $\mathsf{C}^{op}$ to $\mathsf{D}$ can also be regarded as a functor from $\mathsf{C}$ to $\mathsf{D}$ in the following sense.
\begin{defn}
A functor $F:\mathsf{C}^{op}\to\mathsf{D}$ is called a contravariant functor on $\mathsf{C}$ such that it takes each arrow $f:A\to B$ in $\mathsf{C}$ to $F(f):F(B)\to F(A)$ and satisfies that $F(g\circ f)=F(f)\circ F(g)$ for any $f:A\to B$ and $g:B\to C$ in $\mathsf{C}$.
\end{defn}
In contrast with contravariant functor, an ordinary functor is also called a {\em covariant\/} functor. A well-known example of functor is the (covariant) powerset functor ${\mathcal P}:\mathsf{Set}\to\mathsf{Set}$ and the contravariant powerset functor $\breve{\mathcal P}:\mathsf{Set}\to\mathsf{Set}$. For any set $X$, both ${\mathcal P}X$ and $\breve{\mathcal P}X$ are the powerset of $X$ and for any function $f:X\to Y$, ${\mathcal P}f:{\mathcal P}X\to {\mathcal P}Y$ and $\breve{\mathcal P}f:{\mathcal P}Y\to {\mathcal P}X$ are respectively defined by:
\[{\mathcal P}f(U)=f[U]:=\{f(x)\mid x\in U\},\;\mbox{\rm for any}\; U\subseteq X\]
and
\[\breve{\mathcal P}f(V)=f^{-1}[V]:=\{x\in X\mid f(x)\in V\},\;\mbox{\rm for any}\; V\subseteq Y.\]

For fixed categories $\mathsf{C}$ and $\mathsf{D}$, we can consider functors between them as objects of a new category. Then, the arrows between these functors (i.e.\ objects in the new category) are called natural transformations. Formally, we have the following definition.
\begin{defn}
Let $F, G:\mathsf{C}\to\mathsf{D}$ be two functors between categories $\mathsf{C}$ and $\mathsf{D}$. Then, a natural transformation $\vartheta: F\Rightarrow G$ is a family of arrows in $Ar(\mathsf{D})$, indexed by objects in $Ob(\mathsf{C})$, denoted by
\[(\vartheta_C: F(C)\to G(C))_{C\in Ob(\mathsf{C})},\]
such that, for any $f:C\to C'$ in $Ar(\mathsf{C})$, \[\vartheta_{C'}\circ F(f)= G(f)\circ \vartheta_C.\]
In diagram, this means the following commutativity:
\center{
\begin{tikzcd}
F(C)\arrow[r,"\vartheta_C"] \arrow[d,"F(f)"]& G(C) \arrow[d,"G(f)"]\\
F(C')\arrow[r,"\vartheta_{C'}"] & G(C')\\
\end{tikzcd}}
\end{defn}
Given a natural transformation $\vartheta: F\Rightarrow G$, the arrow $\vartheta_C\in Ar(\mathsf{D})$ is called the component of $\vartheta$ at $C$.
\section{Examples of Many-Valued Coalgebraic Modal Logic}
As many-valued coalgebraic modal logic provides an uniform framework for a variety of many-valued modal logics, we instantiate it to some specific examples in this appendix.
\begin{eg}
We can define the crisp Kripke model for many-valued modal logic (\cite{bou2011minimum}) coalgebraically by using the powerset functor $\mathcal P$. Let $\Lambda=\{\Box,\Diamond\}$ be the set of predicate liftings and let $\langle W,\sigma, V\rangle$ be a $\mathcal P$-model. The predicate liftings $\Box,\Diamond: Hom(W,A)\to Hom({\mathcal P}W,A)$ are defined by
\[\Box(f)(X)=\bigwedge_{x\in X}f(x)\]
\[\Diamond(f)(X)=\bigvee_{x\in X}f(x)\]
for any $f:W\to A$ and $X\subseteq W$.  Then, $\sigma:W\to{\mathcal P}W$ corresponds to the functional representation of the binary accessibility relation on $W$ and by definition, the interpretation of modal formulas is
\[\norm{\Box\varphi}_\sigma(w)=\bigwedge_{u\in \sigma(w)}\norm{\varphi}_\sigma(u)\]and
\[\norm{\Diamond\varphi}_\sigma(w)=\bigvee_{u\in \sigma(w)}\norm{\varphi}_\sigma(u).\]
Thus, the coalgebraic framework can accommodate the semantics proposed in \cite{bou2011minimum}.
\end{eg}

\begin{eg}
Let $H$ be the Hom-functor defined in Section~\ref{sec2} and we still consider modalities in $\Lambda=\{\Box,\Diamond\}$. Then, an $H$-model $\langle W,\sigma, V\rangle$ is defined such that $\sigma:W\to Hom(W,A)$ is the functional representation of the $A$-valued accessibility relation on $W$. The predicate liftings $\Box,\Diamond: Hom(W,A)\to Hom(HW,A)$ are defined by
\[\Box(f)(g)=\bigwedge_{x\in W}g(x)\to f(x)\]
\[\Diamond(f)(g)=\bigvee_{x\in W}g(x)\odot f(x)\]
for any $f, g:W\to A$.  Hence, the interpretation of modal formulas is
\[\norm{\Box\varphi}_\sigma(w)=\bigwedge_{u\in W}(\sigma(w)(u)\to \norm{\varphi}_\sigma(u))\]and
\[\norm{\Diamond\varphi}_\sigma(w)=\bigvee_{u\in W}(\sigma(w)(u)\odot\norm{\varphi}_\sigma(u)),\]
which is exactly the same as the Kripke semantics in \cite{bou2011minimum}.
\end{eg}

\begin{eg}
We can characterize the neighborhood semantics in the coalgebraic setting by using the functor $H^2$ (i.e. the composition of the functor $H$ with itself). In an $H^2$-model $\langle W,\sigma, V\rangle$,  $\sigma:W\to Hom(Hom(W,A),A)$ is exactly the $A$-valued neighborhood function defined in \cite{Cintula18}. We only consider a modality $\Box$ because for $A$-valued neighborhood semantics, different modalities are interpreted in the same way but with different neighborhood functions. The predicate lifting $\Box: Hom(W,A)\to Hom(H^2W,A)$ is defined by
\[\Box(f)(N)=N(f)\] for any $f:W\to A$ and $N:(W\to A)\to A$. Hence, the interpretation of $\Box\varphi$ is
\[\norm{\Box\varphi}_\sigma(w)=\sigma(w)(\norm{\varphi}_\sigma),\]
precisely as that given in \cite{Cintula18}. It is easy to extend the framework to deal with multiple modalities at the same time by using the product of functors. For example, we can replace the functor $H^2$ with $H^2\times H^2$ to give semantics for both $\Box$ and $\Diamond$.
\end{eg}

\begin{eg}
The conditional logic based on the selection function semantics\cite{che} has been also presented with the coalgebraic framework in \cite{SchroderP09}. Here, we use an analogous approach to generalize it to many-valued conditional logic. Let us define the selection functor ${\mathcal S}:{\mathsf Set}\to {\mathsf Set}$ that maps a set $X$ to the set $HX\to HX$ of fuzzy selection functions and a function $f:X\to Y$ to ${\mathcal S}f:{\mathcal S}X\to{\mathcal S}Y$ defined by\footnote{We include the definition here only for the sake of completeness. It actually does not play a role in the definition of the semantics.} \[{\mathcal S}f(s)(g)(y)=\bigvee_{x:f(x)=y}s(g\circ f)(x),\]
for any $s:HX\to HX$, $g:Y\to A$, and $y\in Y$. Now, in an ${\mathcal S}$-model $\langle W,\sigma, V\rangle$, $\sigma:W\to {\mathcal S}W$ associates with each possible world $w$ a selection function $\sigma(w):Hom(W,A)\to Hom(W,A)$. Then, we consider a binary modality $\rhd$. To define the predicate lifting for the modality, we first recall the standard definition of the degree of inclusion between two $A$-valued fuzzy sets $f,g\in Hom(W,A)$ as \[f\subseteq g:=\bigwedge_{w\in W}(f(w)\to g(w))\] and then, the predicate lifting $\rhd: Hom(W,A)\times Hom(W,A)\to Hom({\mathcal S}W,A)$ is defined by
\[\rhd(f,g)(s)=(s(f)\subseteq g).\] As a result, the interpretation of the conditional formula is
\[\norm{\varphi\rhd\psi}_\sigma(w)=\sigma(w)(\norm{\varphi}_\sigma)\subseteq\norm{\psi}_\sigma.\]
\end{eg}

\begin{eg}
By using the distribution functor, it is also possible to accommodate probabilistic reasoning about fuzzy events~\cite{Hajek07,hajekGE95} in the framework of many-valued coalgebraic modal logic~\cite{SchroderP11}. To do probability calculation, we assume that the domain of truth values $A$ can be embedding into the unit interval $[0,1]$ so that we can meaningfully combine the probability and truth values by arithmetic operations. The distribution functor ${\mathcal D}:{\mathsf Set}\to {\mathsf Set}$ maps a set $X$ to ${\mathcal D}X:=\{\mu:X\to[0,1]\mid\sum_{x\in X}\mu(x)=1\}$ (i.e., the set of all discrete probability distributions over $X$) and a function $f:X\to Y$ to ${\mathcal D}f:{\mathcal D}X\to{\mathcal D}Y$ such that
\[{\mathcal D}f(\mu)(y)=\sum_{x:f(x)=y}\mu(x)\]
for any $\mu\in{\mathcal D}X$.  Then, for a ${\mathcal D}$-model $\langle W,\sigma, V\rangle$, $\sigma:W\to {\mathcal D}W$ assigns to each possible world a probability distribution over $W$. We exemplify two instances of probability modalities in the coalgebraic framework, $\mathbf P$ and ${\mathbf M}_r$ where $r\in[0,1]$ is a rational number, meaning ``probably'' and ``with probability more than r'' respectively~\cite{SchroderP11}. The predicate liftings of these modalities ${\mathbf P}, {\mathbf M}_r: Hom(W,A)\to Hom({\mathcal D}W,A)$ are defined as follows:
\[{\mathbf P}(f)(\mu)=\sum_{x\in W}f(x)\cdot\mu(x),\]
\[{\mathbf M}_r(f)(\mu)=\bigvee\{\alpha\mid\mu(f_\alpha)>r\},\]
for any $f:W\to A$ and $\mu\in{\mathcal D}W$, where $f_\alpha:=\{x\in W\mid f(x)\geq\alpha\}$ is the $\alpha$-cut of $f$ (regarded as an $A$-valued fuzzy set).  Thus, by definition, the interpretation of probability modal formulas is
\[\norm{{\mathbf P}\varphi}_\sigma(w)=\sum_{x\in W}\norm{\varphi}_\sigma(x)\cdot\sigma(w)(x),\]
\[\norm{{\mathbf M}_r\varphi}_\sigma(w)=\bigvee\{\alpha\mid\sigma(w)((\norm{\varphi}_\sigma)_\alpha)>r\},\] where $\norm{\varphi}_\sigma: W\to A$ is regarded as an $A$-valued fuzzy set.
\end{eg}

\end{document}